\documentclass[journal]{IEEEtran}
\usepackage{amsmath,amsfonts}
\usepackage{algorithmic}
\usepackage{array}
\usepackage{textcomp}
\usepackage{stfloats}
\usepackage{url}
\usepackage{verbatim}
\usepackage{graphicx}
\def\BibTeX{{\rm B\kern-.05em{\sc i\kern-.025em b}\kern-.08em
    T\kern-.1667em\lower.7ex\hbox{E}\kern-.125emX}}
\usepackage{balance}
\usepackage{subfigure}
\usepackage{multirow}
\usepackage{threeparttable}
\usepackage{color}
\usepackage{cite}
\usepackage{flushend}

\begin{document}
\title{Dual Aspect Self-Attention based on Transformer for Remaining Useful Life Prediction}
\author{
	\vskip 1em
	
    Zhizheng Zhang, Wen Song, Qiqiang Li
    \thanks{\copyright  2022 IEEE.  Personal use of this material is permitted.  Permission from IEEE must be obtained for all other uses, in any current or future media, including reprinting/republishing this material for advertising or promotional purposes, creating new collective works, for resale or redistribution to servers or lists, or reuse of any copyrighted component of this work in other works.
    }
    \thanks{This work was supported in part by the National Natural Science Foundation of China under Grant 62102228, in part by the Shandong Provincial Natural Science Foundation under Grant ZR2021QF063, and in part by the Young Scholar Future Plan of Shandong University under Grant No. 62420089964188. \emph{(Corresponding author: Wen Song and Qiqiang Li)}}
	\thanks{Zhizheng Zhang, Wen Song and Qiqiang Li are with the Institute of Marine Science and Technology, Shandong University, China (Email: 202020861@mail.sdu.edu.cn, wensong@email.sdu.edu.cn, qqli@sdu.edu.cn).}
	\thanks{Digital Object Identifier 10.1109/TIM.2022.3160561}
}

\markboth{IEEE Transactions on Instrumentation and Measurement}%
{}

\maketitle

\begin{abstract}
Remaining useful life prediction (RUL) is one of the key technologies of condition-based maintenance, which is important to maintain the reliability and safety of industrial equipments. Massive industrial measurement data has effectively improved the performance of the data-driven based RUL prediction method. While deep learning has achieved great success in RUL prediction, existing methods have difficulties in processing long sequences and extracting information from the sensor and time step aspects. In this paper, we propose Dual Aspect Self-attention based on Transformer (DAST), a novel deep RUL prediction method, which is an encoder-decoder structure purely based on self-attention without any RNN/CNN module. DAST consists of two  encoders, which work in parallel to simultaneously extract features of different sensors and time steps. Solely based on self-attention, the DAST encoders are more effective in processing long data sequences, and are capable of adaptively learning to focus on more important parts of input. Moreover, the parallel feature extraction design avoids mutual influence of information from two aspects. Experiments on two widely used turbofan engines datasets show that our method significantly outperforms the state-of-the-art RUL prediction methods. 
\end{abstract}


\section{Introduction}

\IEEEPARstart{M}{aintenance} management plays a very important role in the operation of modern large mechanical equipment \cite{zhang2016multiobjective}. With the rapid development of modern instruments and measurement technology, it's possible to obtain condition monitoring data from running mechanical equipment \cite{qian2015remaining, zhao2021feature,mao2019predicting}. Condition-based maintenance (CBM) is a maintenance method that requires the measurement of various parameters of the equipment and reflects the actual status of the equipment at any time \cite{cui2019research}. Compared with the traditional preventive maintenance strategy, CBM is more effective in reality due to the utilization of real-time system health information, hence is widely used in the maintenance of modern industrial equipment \cite{2017RUL}. CBM involves predicting the remaining useful life (RUL) and potential faults of the equipment according to the real-time operational status, based on which the maintenance decisions can be made on an as-needed basis according to the prediction information. Obviously, RUL prediction is one of the most critical technologies for effective implementation of CBM. If the RUL of mechanical equipment is predicted according to the current or historical operation information, the time of failure can be accurately known \cite{babu2016deep}. Therefore, RUL prediction is of great importance to researchers in the field of CBM.

In general, RUL prediction methods can be roughly divided into traditional model-based methods, data-driven methods and hybrid methods. The model-based method requires accurate dynamic modeling of mechanical equipment or components to describe the degradation trend of components \cite{park2016model}. However, the structure of modern industrial large-scale equipment is becoming more and more complex, with  miscellaneous nonlinear relationships between various systems and parts. Therefore, it is unrealistic to establish an accurate model. 

The goal of data-driven RUL prediction method is to establish the mapping relationship between RUL and features of the target equipment \cite{qin2020gated}. It does not require extensive expert knowledge and physical modeling for complex mechanical equipment \cite{meng2019review}. In the literature, some traditional machine learning algorithms have been used for RUL prediction, such as support vector regression (SVR) \cite{loutas2013remaining}, random forest (RF) \cite{zhang2016multiobjective}, and extreme learning machine (ELM) \cite{liu2018method}. However,  these methods rely on tedious feature engineering. In contrast, deep learning based methods can automatically extract valuable features from the original CBM data, and achieve much better prediction performance. Consequently, deep learning based RUL prediction methods have a wider range of applicability and have received increasing attention recently \cite{guo2019review}.

The RUL prediction of mechanical equipment is essentially a multivariate time series regression task. The strong temporal and spatial correlation in the condition monitoring signals can be effectively captured by modern deep architectures such as recurrent neural network (RNN) \cite{malhi2011prognosis} and convolution neural network (CNN) \cite{san2020multitask}, which have been widely used in RUL prediction. RNN based methods employ sequence components such as long short-term memory (LSTM) \cite{zheng2017long} and gated recurrent unit (GRU) \cite{chen2019gated} to analyze the signal data sequences. However, due to the existence of recurrent structures in RNN, sequence data needs to pass through each processing unit in turn to extract useful features, which inevitably causes the problem of forgetting important information and is less effective in learning long-term dependencies. CNN based methods normally apply one-dimensional convolution and pooling filters along the time dimension over all sensors to extract feature information \cite{li2018remaining}. However, when processing long time sequences, CNN based methods need to continuously increase the size of convolution kernels to obtain larger time step receptive field which contain more sequence information. That is, the ability of CNN based methods in capturing long-term dependent information in sequence data is also limited.

For RUL prediction, another key issue is that more attention should be paid to the important features that contain more degradation information. Attention mechanism \cite{bahdanau2015neural} is an effective method to learn such dependencies, i.e. weights among different time steps and sensors. Recently, several works attempt to combine the attention mechanism with RNN/CNN based structure to predict RUL \cite{liu2020remaining,xiang2020lstm,song2020distributed,chen2020machine}. However, there are two major shortcomings in these methods. Firstly, the inefficiency of RNN/CNN in capturing long-term dependencies still can not be avoided. Secondly, 
the input data enters the attention modules and RNN/CNN modules sequentially, which 
causes the mutual influence between the extracted feature information, thereby affecting the RUL prediction performance.

Transformer \cite{vaswani2017attention} is a recently proposed sequence modeling architecture. It makes use of the self-attention mechanism to capture the long-term dependencies between elements in a sequence without considering their distance, so that it is less affected by the increase of sequence length compared with traditional methods such as RNN and LSTM. However, for RUL prediction, the vanilla transformer architecture only attention the weights of different time steps and ignores the importance of different sensors in the CBM data stream, which is crucial for the overall prediction performance.

To overcome the above issues, in this paper, we propose a novel deep RUL prediction method named Dual Aspect Self-attention based on Transformer (DAST). In this method, we apply the Transformer architecture \cite{vaswani2017attention} to RUL prediction for the first time, which is an encoder-decoder structure purely based on self-attention, without any RNN/CNN module. DAST consists of two encoders, i.e. the sensor encoder and time step encoder. Each of the two DAST encoders employs the self-attention mechanism to process all CBM sequence data, and automatically learns to pay different attentions to different sensors and time steps. Specifically, the two decoders work \emph{in parallel} in the process of feature extraction, therefore the mutual influence of the two aspect information is avoided. 
The features extracted by the two encoders are fused together and fed into the self-attention based decoder to obtain the RUL prediction.
The main contributions of this paper are summarized as follows:
\begin{enumerate}
	\item We propose a novel end-to-end deep RUL prediction method based on the Transformer architecture. Experiments on two widely used NASA's turbofan engine datasets show that our method significantly outperforms the state-of-the-art RUL prediction methods. 
	\item Based on self-attention, our method is able to automatically pay more attentions to the features that are more important without any domain knowledge, and is more effective in handling long CBM data sequences than RNN/CNN based methods. Our novel dual-aspect design enables extracting features from the sensor and time step dimensions simultaneously, which overcomes the limitation of vanilla Transformer and effectively improves the RUL prediction performance.
	\item The weights of different sensors and time steps learned by our method is intuitive and interpretable to the maintenance personnel, so that they can formulate better maintenance strategies to improve efficiency.
\end{enumerate}

The rest of this paper is organized as follows. The second section introduces the related literature review. The third section introduces the proposed method. In addition, the fourth part also demonstrates the effectiveness and superiority of this method. Finally, the fifth part discusses the advantages compared with the exist methods and the sixth part summarizes the paper.

\section{Related Works}
By modeling the functional relationship between the equipment degradation process and the condition monitoring data, the method based on deep learning can automatically capture the important feature information from the original data to achieve end-to-end prediction. In this section, we briefly review recent deep learning based methods for RUL prediction.

Due to its advantages in processing condition monitoring sequence data, RNN based architecture and its variants (e.g. LSTM and GRU) have been widely used in RUL prediction. Cheng et al. \cite{cheng2020remaining} used the LSTM network prediction model to verify the effectiveness of LSTM compared to RNN. Chen et al. \cite{chen2019gated} proposed a GRU based method for predicting nonlinear deterioration process. Besides RNN, CNN architecture has also been applied to the RUL prediction task. Li et al. \cite{li2018remaining} proposed a RUL prediction method based on deep convolution neural network (DCNN), which directly uses normalized raw data as input and performs convolution operation along the time dimension. Zhu et al. \cite{zhu2018estimation} proposed the multi-scale convolutional neural network (MSCNN) for RUL prediction, which keeps the global and local information synchronously compared to traditional CNN. Some studies attempt to combine clustering and deep learning to improve RUL prediction performance. Javed et al. \cite{javed2015new} proposed a prognostics model using the subtractive-maximum entropy fuzzy clustering to simultaneously predict machine degradation. 
Liu et al. \cite{liu2021prediction} proposed a multi-stage LSTM with clustering for RUL prediction, which first divides data into multiple stages through clustering analysis, and then extract degradation feature information through the LSTM model.

Recently, there are several works employ the attention mechanism to capture the importance of time steps from monitoring data to improve the performance of RNN/CNN based models. For example, Liu et al. \cite{liu2020remaining} proposed a RUL prediction method based on the combination of attention mechanism, GRU and CNN. The features extracted by the attention mechanism are fed into the bidirectional GRU and CNN network to predict RUL. Xiang et al. \cite{xiang2020lstm} proposed a gear RUL prediction method based on LSTM and attention mechanism. Song et al. \cite{song2020distributed} used the distributed attention mechanism and Temporal Convolutional Network (TCN) to predict the engine RUL, which can capture more effective degraded feature information through the attention mechanism. Chen et al. \cite{chen2020machine} proposed to combine the attention mechanism and LSTM with handcrafted features. However, in most of these studies, attention mechanism is used in combination with RNN/CNN architecture, hence the limitation in processing long sequences still exists.

Owing to its effectiveness in modeling long sequences, Transformer has been employed in time-series related tasks recently. Zhou et al. \cite{zhou2021informer} studied the application of Transformer in long sequence time-series prediction and proposed the ProbSparse self-attention mechanism to reduce the time complexity and memory usage. Beltagy et al. \cite{beltagy2020longformer} proposed the Longformer with an attention mechanism that scales linearly with sequence length, making it easy to process long sequence. A recent survey of transformer variants can be found in \cite{lin2021survey}. However, most of existing studies only consider capturing the dependencies between time steps. In the RUL prediction, the weight information between different sensors also has a great influence on the final RUL prediction. Therefore, the above-mentioned research is not suitable for RUL prediction. At present, there are few studies on the application of the Transformer architecture to RUL prediction. 
In this paper, we explore this direction and propose a duel aspect self-attention design to capturing the weight information of different time steps and sensors at the same time, which makes it better to the RUL prediction task.

\begin{figure*}[t]
\centerline{\includegraphics[scale = 0.5]{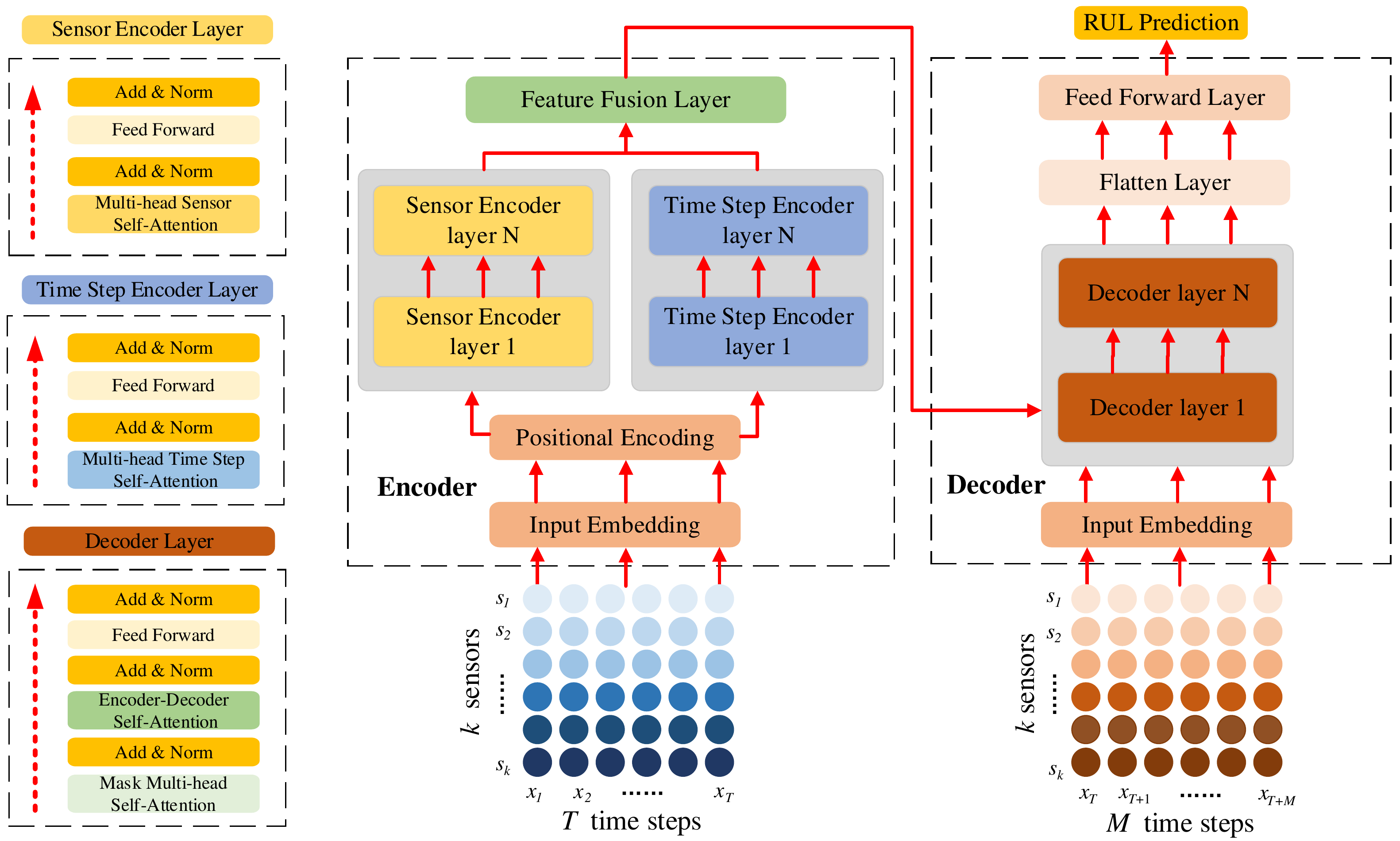}}
\caption{Architecture of DAST.}
\label{DAST}
\end{figure*}

\section{Methodology}
In this section, we first describe the RUL prediction problem, and then present the proposed DAST method in detail, including its architecture and key components.

\subsection{Problem Description}
In this paper, the CBM data collected by sensors during the process of turbofan engine operation are used. 
The RUL prediction problem can be formally defined as follows. The input is $X_t \in \mathbb{R}^k$, $t = (1, 2, \cdots, T)$, where $T$ is the length of time steps and $k$ is the number of sensors. The corresponding output is the predicted RUL $Y_t$ for each time step. Our purpose is to predict real-time RUL by establishing the mapping relationship between RUL and CBM data, expressed as follows:
\begin{equation}{Y_t} = f(X_t),\label{eq1}
\end{equation}
where $X_t$ is the real-time CBM data during operation of the turbofan engine, $f$ is the mapping function, and $Y_t$ is the real-time RUL predicted by $f$. In this paper, we design a Transformer based deep architecture to establish the mapping, which will be detailed in the following subsections.

\subsection{Model Architecture}

From the above description of the RUL prediction problem, it can be seen that the current RUL value of the engine is mainly determined by signals from different sensors at different previous time steps. Therefore, how to make full use of different time steps and sensor information is of great importance to RUL prediction. Moreover, it is intuitive that different sensors and time steps may contain different degradation information, i.e. they may have different degree of importance to the prediction result. In this section, we design a deep architecture based on self-attention that captures the weighted features from both the sensor and time step dimensions, which will be detailed as follows.

Fig. \ref{DAST} shows the architecture of our RUL prediction method, Dual Aspect Self-attention based on Transformer (DAST). DAST follows the encoder-decoder structure in the original Transformer, and consists of three main substructures in the framework: encoder layer (including sensor encoder layer and time step encoder layer), feature fusion layer and decoder layer. Different from the RUL prediction method based on RNN and CNN architecture, DAST captures the long-term dependence information between the inputs and outputs of sequence through self-attention mechanism without considering the distance, so that the importance of each work cycle information will not be reduced due to the increase of time step length. Based on the Transformer architecture, we propose a novel feature extract and fusion approach that is more suitable for the RUL prediction task. It enables learning the weights of different sensors and time steps at the same time, and obtain more valuable feature information of turbofan engine by fusing the feature information of the two parts.

The workflow of our DAST model is as follows. First, it performs feature extraction on CBM data collected by multiple sensors. Specifically, we design a dual aspect encoding mechanism, which applies the sensor encoder and time step encoder that work in parallel to capture the weight features of different sensors and time steps. Both encoders are designed based on multi-head self-attention mechanism. Second, the features extracted from the two aspects are integrated by the feature fusion layer to get a new feature map with importance information of different sensors and time steps. Finally, the fused feature map is sent to the decoder, which adopts the self-attention mechanism to realize the attention of current work cycle information and the previous different time steps and sensors information and outputs the predicted RUL through a fully connected feed-forward network (FFN). Next, we will discuss each of the above mentioned substructures in detail.

\subsection{Encoder of DAST }
The encoder of DAST is mainly composed of an input embedding layer, a positional encoding layer, and multiple sensor encoder layers and time step encoder layers. The input embedding layer maps the input state monitoring data to a vector of ${D_{\text{model}}}$ dimension through a FFN, in order to prepare for the following feature extraction process. The remaining components of encoder is described below.

1) \emph{Positional encoding layer}

As mentioned, our DAST model does not contain structure based on RNN or CNN. Consequently, we need to inject some relative position tokens into the sequence, so that the model can make full use of the position information of the sequence. There are currently a variety of positional encoding methods to choose. In this paper, we use sine and cosine functions of different frequencies\cite{vaswani2017attention}:
\begin{equation}P_t(2k) = sin( {t/1000{0^{2k/{D_{\text{model}}}}}} )\label{eq2}
\end{equation}
\begin{equation}P_t(2k + 1) = cos( {t/1000{0^{2k/{D_{\text{model}}}}}} )\label{eq3}
\end{equation}
where $t$ is the time step and $k$ is the sensor dimension. In this way, $P_t$ has a linear relationship with ${P_{t + l}}$, where $l$ is any fixed time step. This allows the model to easily learn to attend according to the relative positions.

2) \emph{Sensor encoder layer}

A sensor encoder layer mainly includes two sub-layers: a multi-head sensor self-attention layer and a FFN layer. As shown in Fig.\ref{DAST}, there is a residual connection and layer normalization  (Add \& Norm) after each sub-layer. The purpose of residual connection is to alleviate the difficulty of training deep neural network. Layer normalization can accelerate the training process and make the model converge faster by normalizing the layer activation value. Sensor encoder layer uses the multi-head self-attention mechanism \cite{vaswani2017attention} to extract the importance of different sensors along the sensor dimension, therefore it can automatically learn to focus on those sensor features with higher weights without  human experience intervention in the training process. Next, we will introduce the working process of the multi-head sensor self-attention.

\begin{figure}[!t]
\centering
\subfigure[Scaled Dot-Product Self-Attention mechanism.]{
\includegraphics[width = \linewidth]{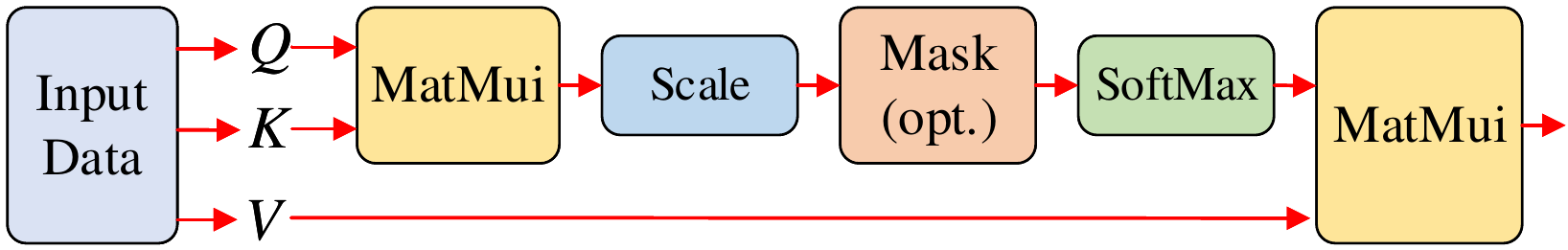}
}

\subfigure[Multi-Head Self-Attention mechanism.]{%
\includegraphics[width = \linewidth]{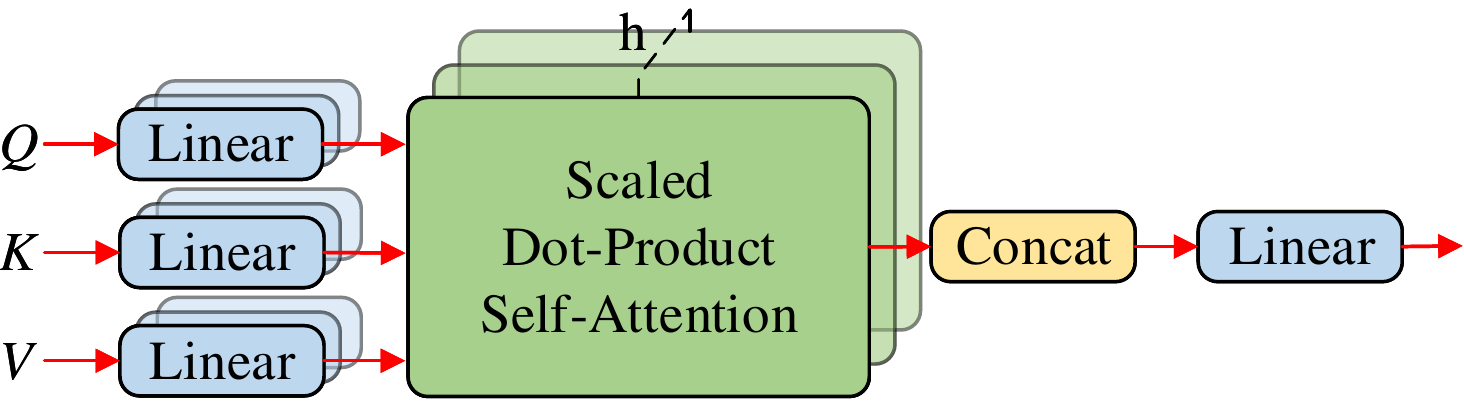}
}
\caption{The process of Multi-Head Self-Attention.}
\label{Attention}
\end{figure}

We define the CBM data collected by the $k$ sensors in a time window of length $T$ as $X = \{X_1, X_2, \cdots, X_T\} \in \mathbb{R}^{d_k \times T}$. We also define $X_s'$ as the data obtained after being processed by the positional encoding layer. The working process of self-attention function is visualized in Fig. \ref{Attention}. First, it generates three matrices (Queries, Keys, Values) by processing the input data using the following computation: 
\begin{equation}Q_s = {X_s'}{W_s^q}, K_s = {X_s'}{W_s^k}, V_s = {X_s'}{W_s^v},\label{eq4}\end{equation}
where ${W_s^q}$, ${W_s^k}$, ${W_s^v}$ are trainable parameters, $Q_s,K_s,V_s \in {\mathbb{R}^{d_k \times D_\text{model}}}$, ${D_{\text{model}}}$ is the input dimension. Then we calculate the dot product of $Q$ and $K$ (scaled by $\sqrt{D_{\text{model}}}$), and apply a softmax function along the sensor dimension to obtain the weights of different sensors in $X_s'$. Therefore, the weight vector of different sensors at time step $t$ is:
\begin{equation}{\alpha_t} = \text{softmax}_{\text{sensors}}( {\frac{{Q_sK_s^T}}{{\sqrt{D_{\text{model}}}}}}) \label{eq7}
\end{equation}
where $\alpha_t = (\alpha_{t,1},{\alpha_{t,2}} \cdots ,{\alpha_{t,k}}), t = (1, 2, \cdots, T)$. Finally, the features of different sensors weighted by the self-attention mechanism is computed as a weighted sum of $V_s$:
\begin{equation}{\text{Attention}}_{\text{sensors}}(Q_s,K_s,V_s) = {\alpha_t}V_s.\label{eq8}
\end{equation}

Here we also adopt the multi-head self-attention mechanism in \cite{vaswani2017attention} to allow the model to jointly attend to information from different representation subspaces at different positions, such that to improve the prediction performance. Fig. \ref{Attention} visualizes multi-head self-attention, which can be expressed as:
\begin{equation}\text{MultiHead}(Q_s,K_s,V_s) = \text{Concat}(\{head_i\}_{i=1}^h){W^s}, \label{eq9}
\end{equation}
where the parameter matrices $W^s \in {\mathbb{R}^{h{D_{\text{model}}} \times D_{\text{model}}}}$, $h$ is the number of heads, and $head_i = \text{Attention}(Q_s,K_s,V_s)_i$.

3) \emph{Time step encoder layer}

As shown in Fig. \ref{DAST}, the time step encoder layer has the same structure as the sensor encoder layer. It mainly includes two sub-layers: a multi-head time step self-attention layer and a FFN layer. The difference is that the time step encoder layer extracts features along the time step dimension, and allows the DAST model to pay attention to the time steps that are more important to the RUL prediction.
The input data of time step encoder is the transpose of $X$ and being processed by the positional encoding layer as $X_t'$. The time step encoder layer first obtains the Queries, Keys, and Values matrices $Q_t,K_t,V_t \in {\mathbb{R}^{T \times D_\text{model}}}$ by Eq. (\ref{eq4}) using trainable parameters ${W_t^q}$, ${W_t^k}$, ${W_t^v}$. Then, the weight vector of different time steps corresponding to the sensor $s$ can be obtained by performing softmax along on the time step dimension:
\begin{equation}{\beta_s} = \text{softmax}_{\text{time\_steps}}( {\frac{{Q_tK_t^T}}{{\sqrt {{D_\text{model}}}}}}) \label{eq10}
\end{equation}
where $\beta_s = (\beta_{s,1},\beta_{s,2} \cdots, \beta_{s,T}), s = (1, 2, \cdots, k)$. The features of different time steps weighted by the self-attention mechanism is obtained as:
\begin{equation}\text{Attention}_{\text{time\_steps}}(Q_t,K_t,V_t) = {\beta_s}V_t.\label{eq11}\end{equation}

Similar to the sensor encoder layer, we also apply multi-head self-attention to the time step encoder layer:
\begin{equation}\text{MultiHead}(Q_t,K_t,V_t) = \text{Concat}(\{head_j\}_{j=1}^h){W^t}, \label{eq9_1}
\end{equation}
where the parameter matrices $W^t \in {\mathbb{R}^{h{D_{\text{model}}} \times D_{\text{model}}}}$, and $head_j = \text{Attention}(Q_t,K_t,V_t)_j$.

4) \emph{Feature Fusion Layer}

\begin{figure}[!t]
\centerline{\includegraphics[width = \linewidth]{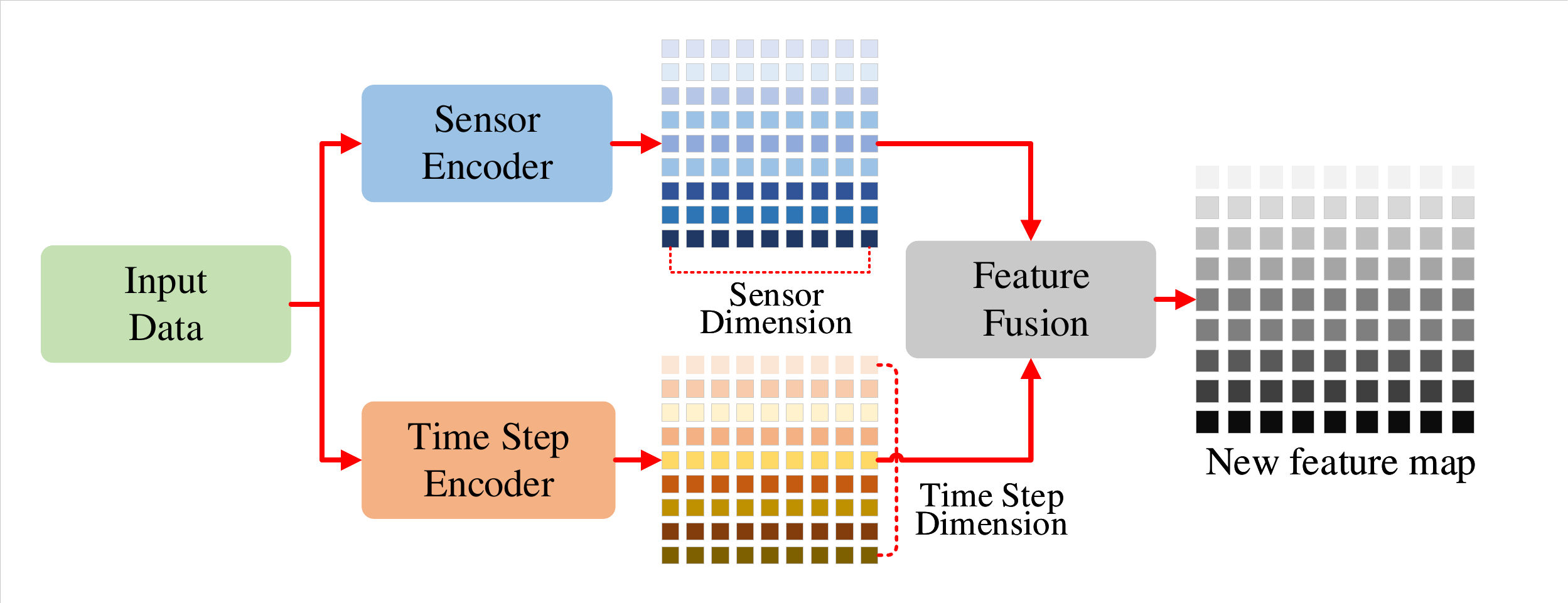}}
\caption{The process of feature fusion.}
\label{feature fusion process}
\end{figure}

After extracting features from the time step and sensor dimension of CBM data, DAST performs feature fusion to integrate information from the two aspects. As shown in Fig. \ref{feature fusion process}, the feature fusion layer combines the time step and sensor features to form a new feature map. We denote the features extracted from the sensor encoder and time step encoder as ${F_s} \in \mathbb{R}^{d_k \times D_{\text{model}}}$ and ${F_t} \in \mathbb{R}^{T \times D_{\text{model}}}$, respectively. DAST performs feature fusion using the following computation:
\begin{equation}{F_r} = \text{Concat}\left({{F_s},{F_t}} \right){W^f}, \label{eq12}\end{equation}
where the trainable parameter matrices $W^f \in \mathbb{R}^{(d_k+T) \times D_{\text{model}}}$, which can make the model to capture feature information from both $F_s$ and $F_t$.

In DAST, both the sensor encoder and time step encoder are constructed by stacking multiple identical sensor or time step encoder layers. Here for convenience, we use the same number of stacks $N$ for both encoders, but in general they could be different. The hyperparamter $N$ could affect the feature extraction capability of DAST. We empirically tune $N$ in the experiments to obtain good performance. We would like to note that unlike previous works \cite{liu2020remaining,xiang2020lstm,song2020distributed,chen2020machine}, in DAST, features of the sensor dimension and time step dimension are extracted simultaneously, since the time step encoder and sensor encoder are arranged in parallel. This design effectively avoids the mutual influence of information from the two aspects, which helps to improve the performance of RUL prediction. At the same time, there is no RNN/ CNN module in DAST, which is purely based on self-attention mechanism to process the long-term dependence information.
We will show the advantage of our design in the experiments.

\subsection{Decoder of DAST}
The decoder of DAST is designed in a similar way as in the original Transformer \cite{vaswani2017attention}. As shown in Fig. \ref{DAST}, the decoder consists of an input embedding layer, multiple identical decoder layers, a flatten Layer and a FFN layer. A decoder layer mainly includes two multi-head self-attention sub-layers, including: mask multi-head self-attention and encoder-decoder multi-head self-attention sublayer. The encoder-decoder multi-head self-attention will receive Keys and Values from the output of the encoder, while Queries are from the output of the previous layer of the decoder. Therefore, the weight features of different sensors and time steps extracted in the encoder part are analyzed on the decoder, which realizes the attention of the current working cycle information and the previous different time steps and sensors information and finally outputs the predicted RUL through the FFN layer. In order to ensure that prediction of a time sequence data point will only depend on previous data points, the masked multi-head self-attention is applied in the self-attention computed by setting the corresponding dot products to $- \infty$. Such mask mechanism can ensure that model only apply attention to the data points before the target data. In other words, the self-attention mechanism will only attention on data ${x_{T - 1}}$ and previous data when we predict the RUL of ${x_{T}}$. In this paper, we follow previous works in RUL prediction and apply the rolling prediction.

\section{Experiments}
In this section, we introduce the experimental dataset, related experimental settings, and conduct experiments on two widely used turbofan engine datasets to evaluate the effectiveness of DAST compared to state-of-the-art RUL prediction methods, and to validate the advantage of the DAST design.

\subsection{Datasets}
In this paper, we mainly adopt the widely used C-MAPSS (Commercial Modular Aero Propulsion System Simulation) dataset \cite{saxena2008damage} to evaluate our method. The C-MAPSS dataset contains four different sub-datasets. As shown in Table \ref{dataset}, FD001, FD002, FD003, and FD004 have different numbers of operating conditions and fault modes. The F002 and F004 datasets have more complex operating conditions and fault modes, so RUL is more difficult to predict than the F001 and F003 datasets. Among the 21 sensors in the C-MAPSS dataset (indexed from 1 to 21), sensors 1, 5, 6, 10, 16, 18 and 19 always have constant values during the run-to-failure experiments, meaning that data from these sensors cannot characterize the degradation process of the engine. Therefore, we remove these sensor data series and use the data of the remaining 14 sensors for RUL prediction.
Another problem we considered is that in actual situation, the degradation process of turbofan engine in the early stage can be ignored, that is to say, the engine RUL should be kept constant in the early stage. Therefore, we follow Zheng et al.\cite{zheng2017long} and limit the engine RUL from start to degradation to $\text{RUL}_{\text{max}}$, and the linear degradation of turbofan engine occurs after $\text{RUL}_{\text{max}}$. In this work, $\text{RUL}_{\text{max}}$ is set to 125. Besides C-MAPSS, we also test our method on the PHM 2008 dataset \cite{PHM2008}. This dataset was widely used for RUL evaluations, which has the same data structure with  C-MAPSS dataset. The details of training and test set are shown in Table \ref{dataset}. Each data set contains a training set and a test set. The training set contains data for each time step in the complete run-to-failure process. 
In the test set, the engine is randomly stopped before the failure occurrence, and the aim is to predict the true RUL of the last time step.

\begin{table}[!t]
\centering
\setlength\tabcolsep{4pt}
\caption{Description of the Datasets}
\label{dataset}
\begin{tabular}{lccccc}
\hline\hline
\multirow{2}{*}{Dataset} & \multicolumn{4}{c}{C-MAPSS}  & \multirow{2}{*}{PHM-2008} \\ \cline{2-5}
 & FD001 & FD002 & FD003 & FD004 &   \\ \hline
Training engines    & 100   & 260   & 100   & 249   & 218 \\ 
Testing engines & 100   & 259   & 100   & 248   & 218 \\ 
Operating conditions & 1     & 6     & 1     & 6     & 6    \\ 
Fault modes        & 1     & 1     & 2     & 2     & 2 \\ 
Training set size & 20631 & 53759 & 24720 & 61249 & 45918 \\
Test set size & 100 & 259   & 100   & 248   & 218 \\ \hline\hline
\end{tabular}
\end{table}

\begin{figure}[!t]
\centerline{\includegraphics[width = \linewidth]{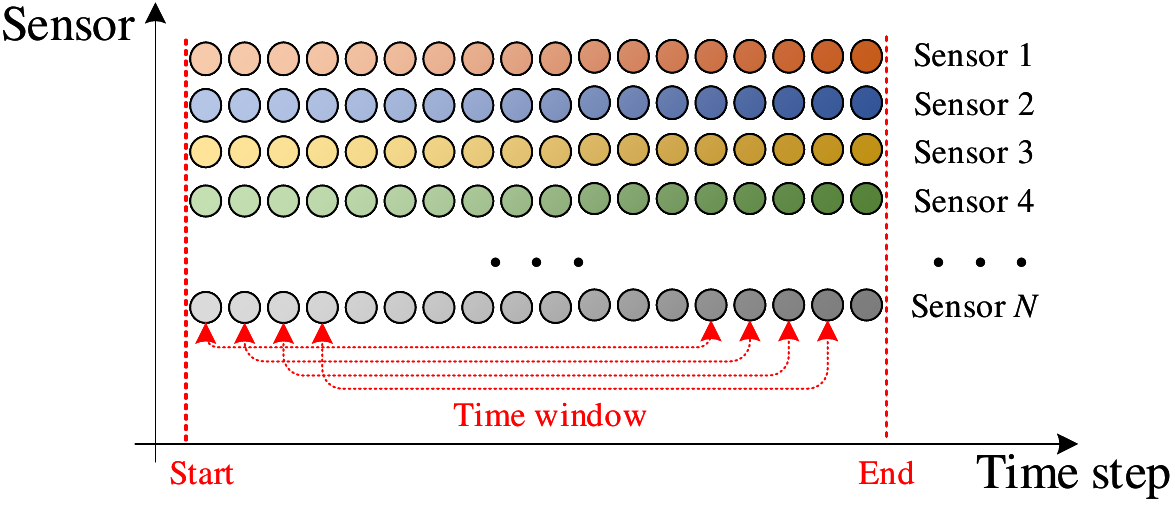}}
\caption{Sliding window processing.}
\label{Sliding window}
\end{figure}

\subsection{Experimental Setting}

\subsubsection{Data preprocessing} We perform the following data preprocessing procedures.

\textbf{Normalization.} 
Data from different sensors have various units and scales, which could affect the accuracy of RUL prediction \cite{li2020remaining}. Hence, we use the min-max scaler method to normalize sensor data. Specifically, for the CBM data $X_i = \{{X_1},{X_2}, \cdots ,{X_T}\}$, we normalize it as follows:
\begin{equation}\widetilde{X}_i = \frac{{{X_i} - {\rm{min}}\left( {{X_i}} \right)}}{{\max \left( {{X_i}} \right) - {\rm{min}}\left({{X_i}} \right)}}, \label{eq13}\end{equation}
where $\widetilde{X}_i$ is the normalized data, ${\rm{max}}\left({{X_i}} \right)$ and ${\rm{min}}\left({{X_i}} \right)$ denote the maximum and minimum of $X_i$.

\textbf{Sliding window processing.} A sliding window is often used for data segmentation preprocessing in order to make the model get valuable information from multivariate time series as much as possible. A simple example of sliding time window processing is shown in Fig. \ref{Sliding window}. $T_w$ is the size of time window and the sliding stride is set to one. The RUL of the last data point in a time window serves as the RUL of that window. We will discuss the impact of time window size on the model prediction performance later.

\textbf{Statistical features.}
The mean value and regression coefficient estimates \cite{song2020distributed} of the sequence data are two explicit numerical features often used in time sequence data, which can provide useful sequence statistics. In this paper, we add these two parts of feature information to the sequence.

\subsubsection{Evaluation metrics}

We use two commonly adopted performance indicators to verify the effectiveness of our method. One is the well-known metric mean square error (RMSE), and the other is the Score metric \cite{saxena2008damage} defined as follows:
\begin{equation}{\rm{Score}} = \left\{ {\begin{array}{*{20}{c}}
{\mathop \sum \limits_{i = 1}^n {e^{ - \frac{{{y_{i}^{'} - y_{i}}}}{{13}}}} - 1,\;\;for\; y_{i}^{'} - y_{i} < 0}\\
{\mathop \sum \limits_{i = 1}^n {e^{\frac{{{y_{i}^{'} - y_{i}}}}{{10}}}} - 1,\;\;for\; y_{i}^{'} - y_{i} \ge 0}
\end{array}} \right.\label{eq14}\end{equation}
where $n$ is the number of testing samples, $y_i'$ is the predicted RUL and $y_i$ is the true RUL for the $i$th sample. Compared to RMSE, the Score metric punishes more when the predicted RUL is larger than the true RUL. This is reasonable because in practice, such ``optimistic" prediction will cause more serious impact. For both RMSE and Score, the lower the value, the better the prediction accuracy.

\subsubsection{Hyperparameters and implementation details}
To determine the structure of our DAST model, we perform grid search to find the best model configuration. The resulting structural parameters are listed in Table \ref{table:DAST_param}. The sliding time window length is set to 40 on the F001 and F003 datasets, and 60 on the F002 and F004 datasets, and we will discuss the impact of the window length later. For training, we use the rectified Adam optimizer and set the epoch to 100. We use RMSE as the training loss function. The learning rate is set to 0.001 and the batch size is set to 256. We apply dropout for each encoder and decoder layer, and the dropout rate is set to 0.2. All experiments are performed on a windows 10 workstation, which is equipped with 64GB RAM and an Intel 9900K CPU. Our code in Pytorch is available at https://github.com/Zzzsdu/DAST.

\begin{table}[!t]
\centering
\caption{Structural Parameters of DAST}
\label{table:DAST_param}
\begin{tabular}{ccc}
\hline\hline
Components  & Layers & Parameters \\ \hline
Input  & \multirow{2}{*}{Fully connected layer}    & Hidden units: 64\\ 
embedding & & Activation: Linear \\ \hline
\multirow{4}{*}{Encoder}  & \multirow{2}{*}{Sensor encoder layer}    & Encoder blocks N = 2 \\ 
 & & Self-attention heads H = 4 \\ \cline{2-3}
 & \multirow{2}{*}{Time step encoder layer} & Encoder blocks N = 2 \\ 
 & & Self-attention heads H = 4 \\\hline
\multirow{2}{*}{Decoder}  & \multirow{2}{*}{Decoder layer }   & Decoder blocks N = 1\\     
 & & Self-attention heads H = 4   \\ \hline
\multirow{4}{*}{Output}  & \multirow{2}{*}{Fully connected layer}    & Hidden units: 64 \\ 
 & & Activation: (ReLU) \\ \cline{2-3}
 & \multirow{2}{*}{Output layer} & Hidden units: 1\\ 
 & & Activation: Linear \\ \hline\hline
\end{tabular}
\end{table}

\begin{table*}[h]
\centering
\begin{threeparttable}
\setlength\tabcolsep{4pt}
\caption{RMSE comparison with state-of-the-art methods}
\label{table:RMSE}
\begin{tabular}{cccccccccc}
\hline\hline
Dataset    & BiLSTM\cite{wang2018remaining}  & DCNN\cite{li2018remaining}  & ELSTMNN\cite{cheng2020remaining}  &Kong et al.\cite{kong2019convolution} &Chen et al.\cite{chen2020machine} &DATCN\cite{song2020distributed} &DARNN\cite{zeng2021deep}  & AGCNN\cite{liu2020remaining} & DAST     \\ \hline

F001       & 13.65   & 12.61 & 18.22 &16.13 &14.53 &11.78* & 12.04 & 12.42 & \textbf{11.43}  \\ 

F002         & 23.18   & 22.36 & /  &20.46 &/ &16.95* & 19.24 & 19.43 & \textbf{15.25}  \\ 

F003          & 13.74   & 12.64 & 23.21  &17.12 &/ &11.56 & \textbf{10.18}* & 13.39 & 11.32    \\ 

F004         & 24.86   & 23.31 & / &23.26 &27.08 &18.23 & \textbf{18.02*} & 21.50 & 18.36  \\ \hline

Average     & 18.85  & 17.73 & / &19.24 &/ &14.63* &14.87 &16.68 & \textbf{14.09}  \\ \hline\hline
\end{tabular}
\begin{tablenotes}
\footnotesize
\item{* means the best result in the baseline methods, \textbf{bold} means the best result in all methods.}
\end{tablenotes}
\end{threeparttable}
\end{table*}

\begin{table*}
\centering
\begin{threeparttable}
\setlength\tabcolsep{4pt}
\caption{Score comparison with state-of-the-art methods}
\label{table:score}
\begin{tabular}{cccccccccc}
\hline\hline
Dataset      & BiLSTM\cite{wang2018remaining}   & DCNN\cite{li2018remaining} & ELSTMNN\cite{cheng2020remaining}  &Kong et al. \cite{kong2019convolution}  &Chen et al.\cite{chen2020machine} &DATCN\cite{song2020distributed} &DARNN\cite{zeng2021deep}  & AGCNN\cite{liu2020remaining}  & DAST        \\ \hline
F001         & 295      & 273.7  & 571 &303 &322.44 &229.48 &261.95 & 225.51* & \textbf{203.15}    \\ 
F002            & 4130    & 10412 & /  &3440 &/ &1842.38 &933.58* & 1492   & \textbf{924.96}   \\ 
F003           & 317      & 284.1 & 839  &1420 &/ &257.11 & 247.85 & 227.09* & \textbf{154.92}   \\ 
F004            & 5430     & 12466 & / &4630 &5649.14 &2317.32* & 2587.44 & 3392   & \textbf{1490.72}  \\ \hline
Average     & 2543  & 5858.95 & / &2448.25 &/ &1161.57 & 1007.71* & 1334.15 & \textbf{693.43}  \\ \hline\hline
\end{tabular}
\begin{tablenotes}
\footnotesize
\item{* means the best result in the baseline methods, \textbf{bold} means the best result in all methods.}
\end{tablenotes}
\end{threeparttable}
\end{table*}

\subsection{Comparison with Other Methods}

Here we compare the performance of DAST with state-of-the-art deep learning based RUL prediction methods. The baselines are from three categories: 1) pure RNN/CNN based methods \cite{wang2018remaining,li2018remaining,cheng2020remaining}, 2) RUL prediction methods based on the combination of RNN/CNN architecture and the attention mechanism \cite{zeng2021deep,liu2020remaining,chen2020machine,song2020distributed}, and 3) health indicators based methods \cite{kong2019convolution,malhotra2016multi,nguyen2021automated}, which are recently proposed to combine device health indicators and deep learning models to improve prediction performance.
To mitigate the impact of randomness,  we repeat all prediction experiments 10 times, and report the average performances throughout the paper. 

\subsubsection{Comparison on the C-MAPSS dataset} 
We first compare our method with the above baselines on the C-MAPSS datset. As shown in Table \ref{table:RMSE} and Table \ref{table:score}, our method consistently outperforms all comparing ones with the smallest value in both RMSE and Score on average, showing that DAST can produce more accurate RUL prediction. More importantly, the improvement of DAST against the best results of existing methods tends to be larger on the two harder datasets F002 and F004. Specifically, the corresponding RMSE of DAST is reduced by 10.02\% on F002 and the Score is reduced by 35.67\% on F004 compared with state of the art. Another key observation is that, the improvement of DAST is much more prominent in terms of Score, showing its advantage in this practical metric. Our method also outperforms recent health indicators construction based methods \cite{kong2019convolution,malhotra2016multi,nguyen2021automated} which requires experienced feature engineering process to generate the representative health indicators. We can see from Table \ref{table:RMSE} and \ref{table:score} that results of DAST are better than \cite{kong2019convolution} on all the four sub-datasets.
For \cite{malhotra2016multi} and \cite{nguyen2021automated} which only reported results on F001, the best result is 12.80 for RMSE and 256 for Score, which are worse than our prediction results in Table \ref{table:RMSE} and \ref{table:score}. To summarize, the above results and discussions show that the proposed DAST model has good ability in modeling complex multivariate time series data and good application prospect in practical RUL prediction.

\begin{figure}[!t]
\centering
\subfigure[F001 Testing Engine unit24]{
\includegraphics[width=0.48\linewidth]{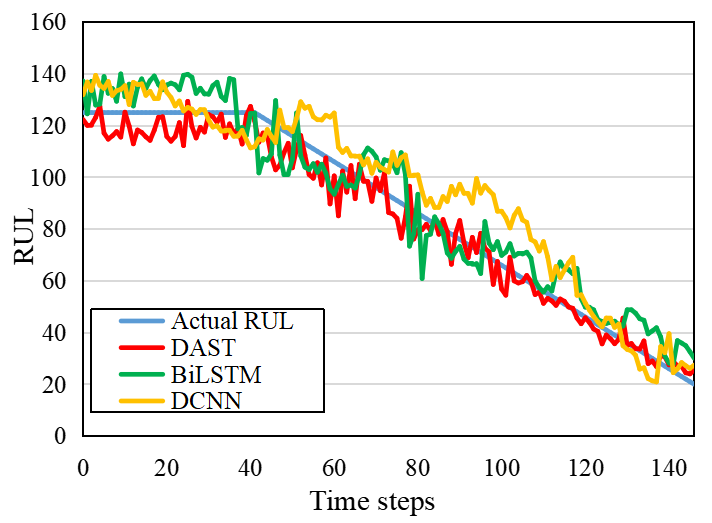}
}\subfigure[F002 Testing Engine unit80]{%
\includegraphics[width=0.48\linewidth]{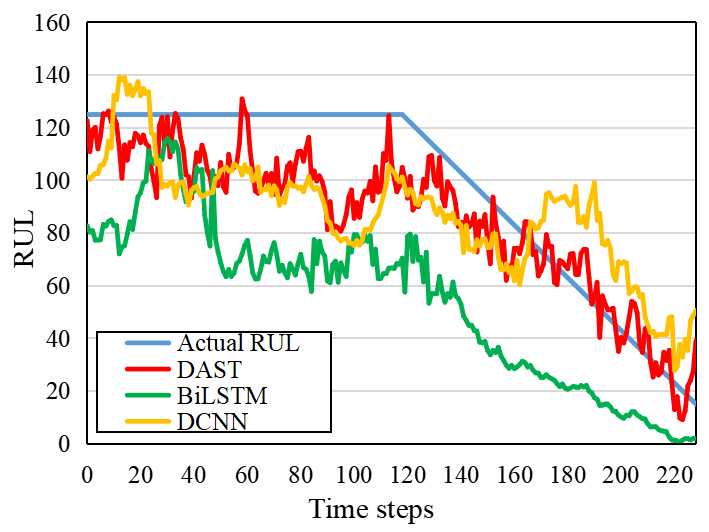}
}
\subfigure[F003 Testing Engine unit99]{%
\includegraphics[width=0.48\linewidth]{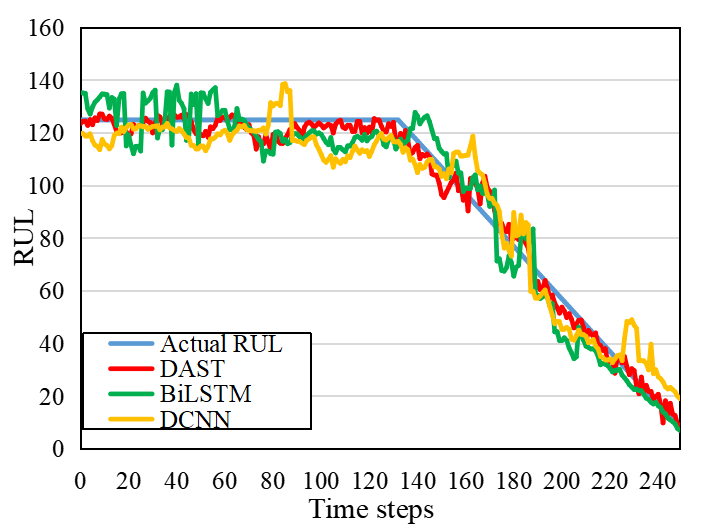}
}\subfigure[F004 Testing Engine unit8]{%
\includegraphics[width=0.48\linewidth]{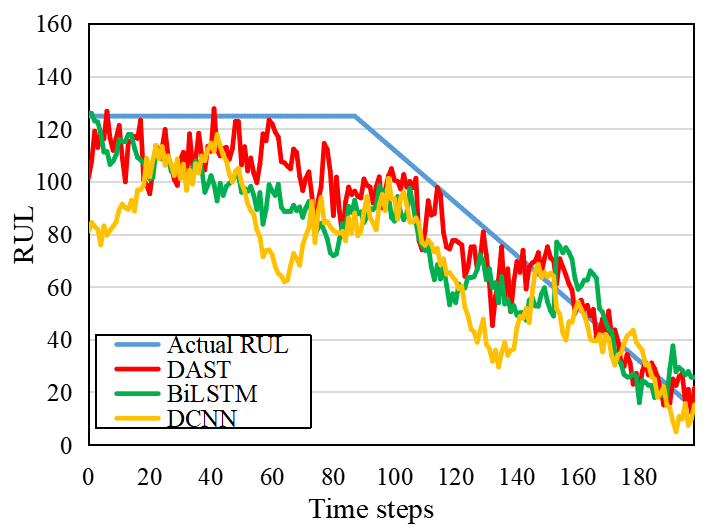}
}
\caption{RUL prediction results on four test engine units.}
\label{RUL prediction results}
\end{figure}

\subsubsection{Detailed analysis of the prediction results} 
Next, we perform more fine-grained analysis on the prediction results. First, to visually analyze the RUL predicted by our DAST model, we randomly select an engine unit from each of four C-MAPSS datasets, and compare the predicted RUL with the actual RUL of engine. We also compare two representative RNN and CNN based methods, i.e. BiLSTM \cite{wang2018remaining} and DCNN \cite{li2018remaining}. All results are plotted in Fig. \ref{RUL prediction results}. We can see that the predicted RUL of our method has a similar trajectory to the real RUL and is clearly better than those of BiLSTM and DCNN, showing the effectiveness of DAST in capturing degradation information. Moreover, most of the RUL values predicted by DAST are close to or smaller than the actual RUL, which is desirable because overestimation of RUL could cause more serious consequences than underestimation. This shows why the improvement of our method in Score reported in Table \ref{table:score} is more significant. Compared to the later stages, the prediction error tends to be larger in the early stages. This is because when the engine begins to enter the degradation stage, the CBM data contains more degradation information, which makes the later predictions more precise \cite{liu2020remaining}. We can also see that compared with the other two datasets, F002 and F004 are more difficult to predict, since they contain more complicated operating conditions and fault modes.

\begin{figure}[!t]
\centering
\subfigure[RMSE comparison of BiLSTM]{
\includegraphics[width=0.48\linewidth]{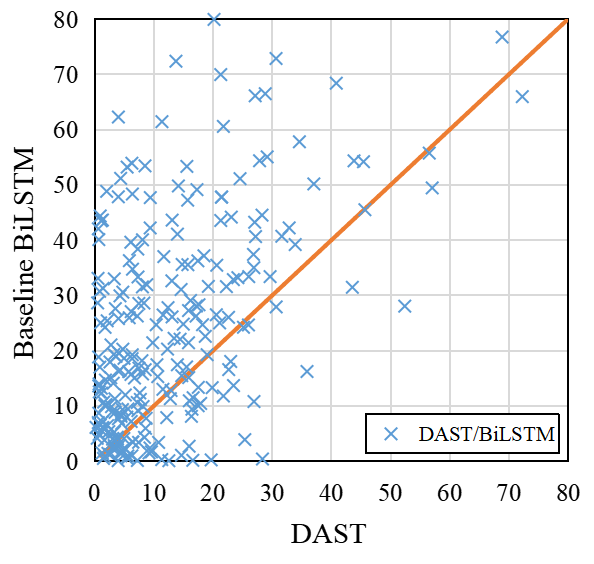} 
}\subfigure[RMSE comparison of DCNN]{%
\includegraphics[width=0.48\linewidth]{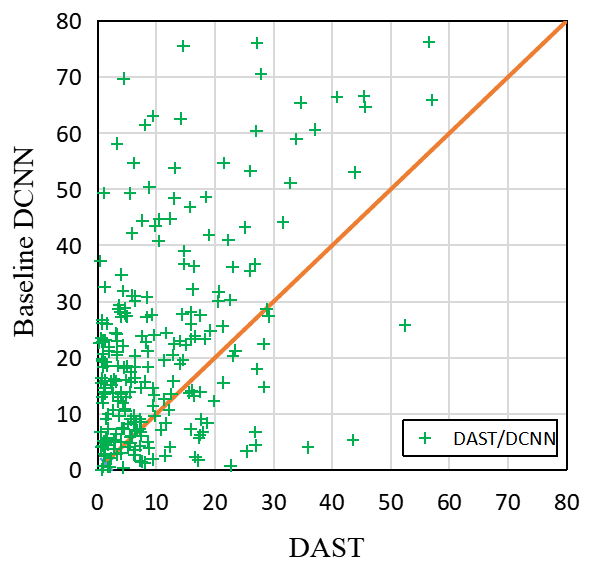}
}
\subfigure[Score comparison of BiLSTM]{%
\includegraphics[width=0.48\linewidth]{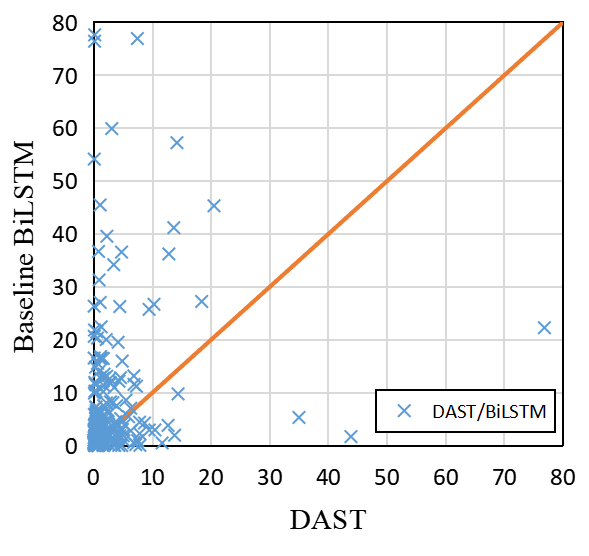}
}\subfigure[Score comparison of DCNN]{%
\includegraphics[width=0.48\linewidth]{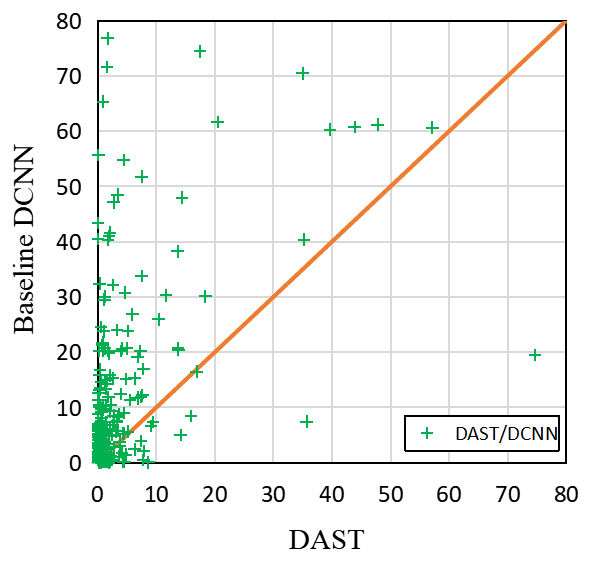}
}
\caption{Pairwise comparison of DAST with BiLSTM and DCNN on F002.}
\label{RMSE and Score 0f F002}
\end{figure}

Then, we conduct pairwise comparison to analyze the performance of DAST against BiLSTM and DCNN. We take F002 dataset as an example, for which we plot the prediction results of DAST and the comparing methods as a pair for all the 259 engines in Fig. \ref{RMSE and Score 0f F002}. Note that points above the diagonal (orange line) indicate that the RMSE/Score values of BiLSTM/DCNN are higher than DAST. We can observe that most of the points distribute above the diagonal, showing that DAST gives better prediction for most engines. In fact, for the 259 engines, DAST has smaller RMSE and Score than BiLSTM on 75.6\% and 72.6\% engines, and also outperforms DCNN on 78.3\% and 81.6\% engines, respectively. 

We further compare the training and testing time efficiency of DAST with the two representative RNN/CNN based methods BiLSTM and DCNN in the same experimental environment, taking F001 dataset as an example. As shown in Table \ref{table:time consumption}, the training time of DAST is higher than that of the CNN based method DCNN. Considering that model training is only required once and offline, this amount of time for training is still acceptable. The testing time of DAST for all the 100 engines is only 0.03s on a CPU, which means that the testing time for one engines is 0.3ms. Hence, the proposed method can meet the requirement of real-time RUL prediction.

\begin{table}[!t]
\centering
\caption{Training and Testing Time Comparison}
\label{table:time consumption}
\begin{tabular}{cccc}
\hline\hline
Method     & DAST    & DCNN\cite{li2018remaining}   & BiLSTM\cite{wang2018remaining}  \\ \hline
Training time (s) & 1229.74 & 715.35 & 1269.78  \\ 
Testing time (s)  & 0.030   & 0.014  & 0.042    \\ 
RMSE          & 11.43   & 12.61  & 13.65    \\ 
Score         & 203.15  & 273.70 & 295.00   \\ 
\hline\hline
\end{tabular}
\end{table}

\subsubsection{Comparison on the PHM 2008 dataset}
For this experiment, we apply the same DAST model as in the C-MAPSS experiments, with sliding time window size $T_w=60$. Because the PHM 2008 dataset does not contain actual RUL labels, we need to upload the prediction results to the NASA data repository website\footnote{https://ti.arc.nasa.gov/tech/dash/groups/pcoe/prognostic-data-repository/} to get the Score metric for evaluation. In Table \ref{table:PHM}, we list the performance of DAST and several recent methods. As we can see, DAST significantly outperforms all the baselines. Compared to the most competitive method in \cite{chen2020machine}, DAST achieves an improvement of 46.7\%, showing its effectiveness in this dataset.

\begin{table}[!t]
\centering
\caption{Results on the PHM 2008 dataset}
\label{table:PHM}
\begin{tabular}{cc}
\hline\hline
Method    & Score  \\ \hline
MLP\cite{sateesh2016deep}         & 3212   \\ 
SVR\cite{sateesh2016deep}         & 15886  \\ 
RVR\cite{sateesh2016deep}         & 8242   \\ 
CNN\cite{sateesh2016deep}         & 2056   \\ 
Deep LSTM\cite{zheng2017long}   & 1862   \\ 
Chen et al.\cite{chen2020machine} & 1584*   \\ 
DAST        & \textbf{845}    \\ \hline
Imp         & 46.7\% \\ 
\hline\hline
\end{tabular}
\end{table}

\subsection{Analysis of DAST}

\subsubsection{Impact of the sliding time windows size} 
\begin{figure}[!t]
\centering
\subfigure{\includegraphics[scale = 0.31]{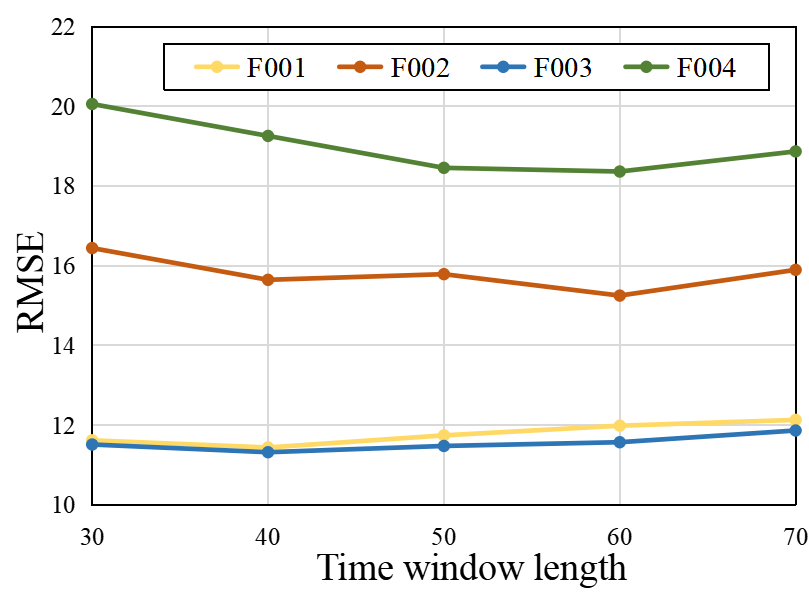}}
\subfigure{\includegraphics[scale = 0.31]{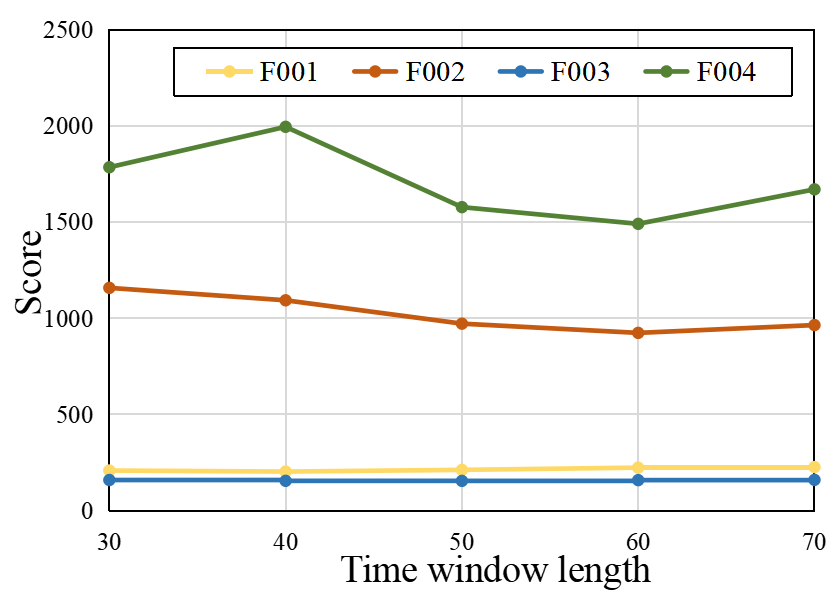}}
\caption{Performance of DAST with different time window sizes on the  C-MAPSS dataset.}
\label{different window sizes}
\end{figure}

The multivariate time series data contains data with different RUL information, so it is necessary to choose a reasonable time window size before training. To verify the influence of this parameter, we conduct several groups of comparative experiments using the four sub-datasets in C-MAPSS, by setting $T_w$ to 30, 40, ..., 70. The results are plotted in Fig. \ref{different window sizes}. We can observe from Fig. \ref{different window sizes} that the RMSE and Score values are the smallest when the time window length is 40 on the F001 and F003 datasets, while for F002 and F004 the two metrics are the best when $T_w=60$. This is because the F002 and F004 datasets have more complex operating conditions and failure modes than the F001 and F003 datasets, so larger time window sizes could contain more degradation information.

\begin{table}[!t]
\centering
\caption{Ablation study on F002}
\label{table:ablation}
\begin{tabular}{lccc}
\hline\hline
Method & Metric & Mean & STD \\ \hline
\multirow{2}{3.7cm}{DAST w/o sensor encoder (vanilla Transformer) }  & RMSE & 16.47 & 0.39 \\
 & Score & 1638.36 & 132.64 \\ \hline
\multirow{2}{*}{DAST w/o time step encoder} & RMSE & 16.11 & 0.35 \\
 & Score & 1476.73 & 119.82 \\ \hline
\multirow{2}{3.7cm}{DAST with sensor and time step encoder arranged in series} & RMSE & 15.86 & 0.21 \\
 & Score & 1367.61 & 49.67 \\ \hline 
\multirow{2}{*}{DAST} & RMSE & \textbf{15.25} & 0.24 \\
 & Score & \textbf{924.96} & 47.67 \\ \hline\hline
\end{tabular}
\end{table}

\subsubsection{Ablation study of DAST} 
Here we evaluate the effectiveness of some internal components of our method using ablation study. More specifically, we evaluate the two key components, i.e. the sensor encoder and time step encoder, as well as the parallel feature extraction design. To this end, we perform three experiments in this part: DAST without sensor encoder, DAST without time step encoder, and DAST with sensor encoder and time step encoder arranged in series. Note that DAST without sensor encoder is essentially the vanilla Transformer. For the third experiment, the input data first pass through the sensor encoder, the output of which is treated as input to the time step encoder. We take the F002 dataset as an example, and the  experimental results are shown in TABLE \ref{table:ablation}. The experimental results show that the prediction performance based DAST is superior than the original Transformer. Moreover, when any of the sensor or time step encoder is removed, the performance significantly drops, showing that information from both aspects are valuable for RUL prediction. DAST also outperforms the version that extracts features in series, which verifies the effectiveness of our parallel design in alleviating the mutual influence between information from the two aspects. 

\subsubsection{Visualization of the learned weights} 
\begin{figure}[!t]
\centering
\subfigure[The average weights of sensors]{\includegraphics[width=\linewidth]{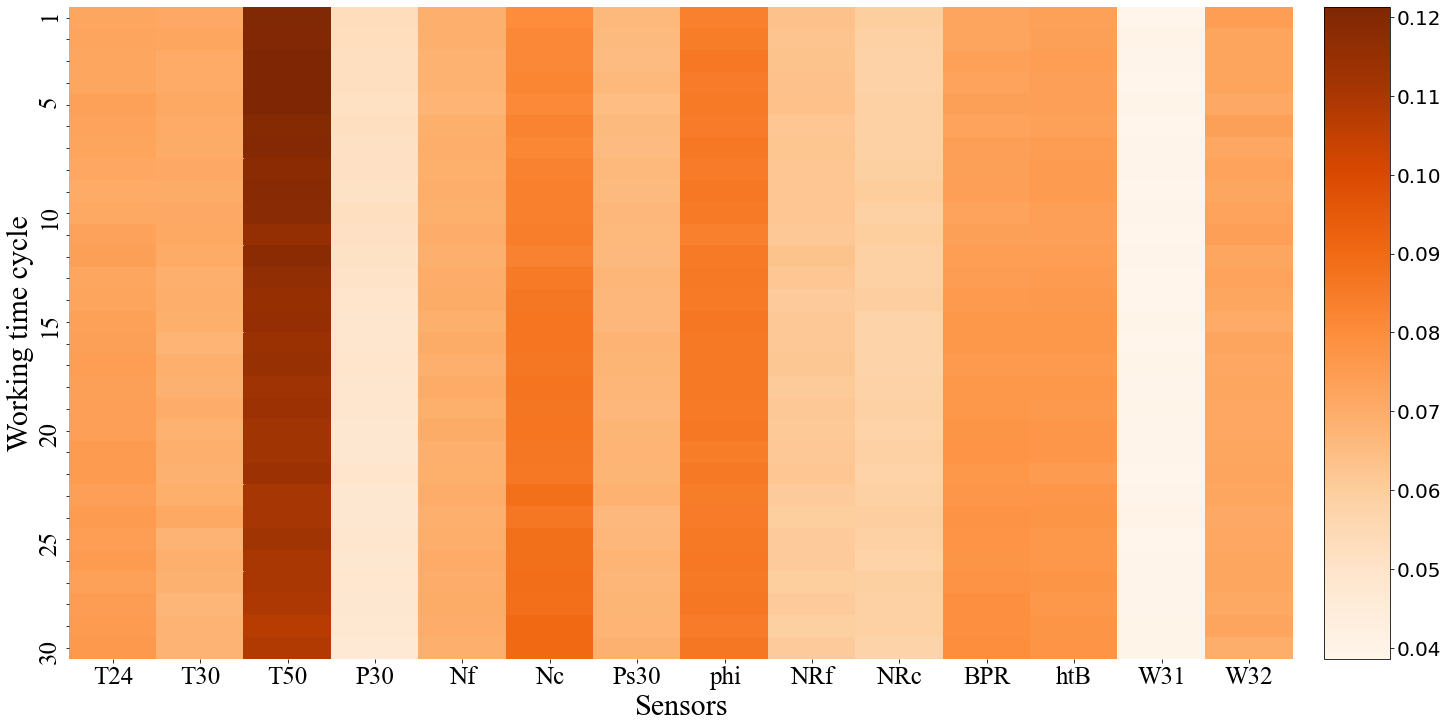}}
\subfigure[The average weights of time steps]{\includegraphics[width=\linewidth]{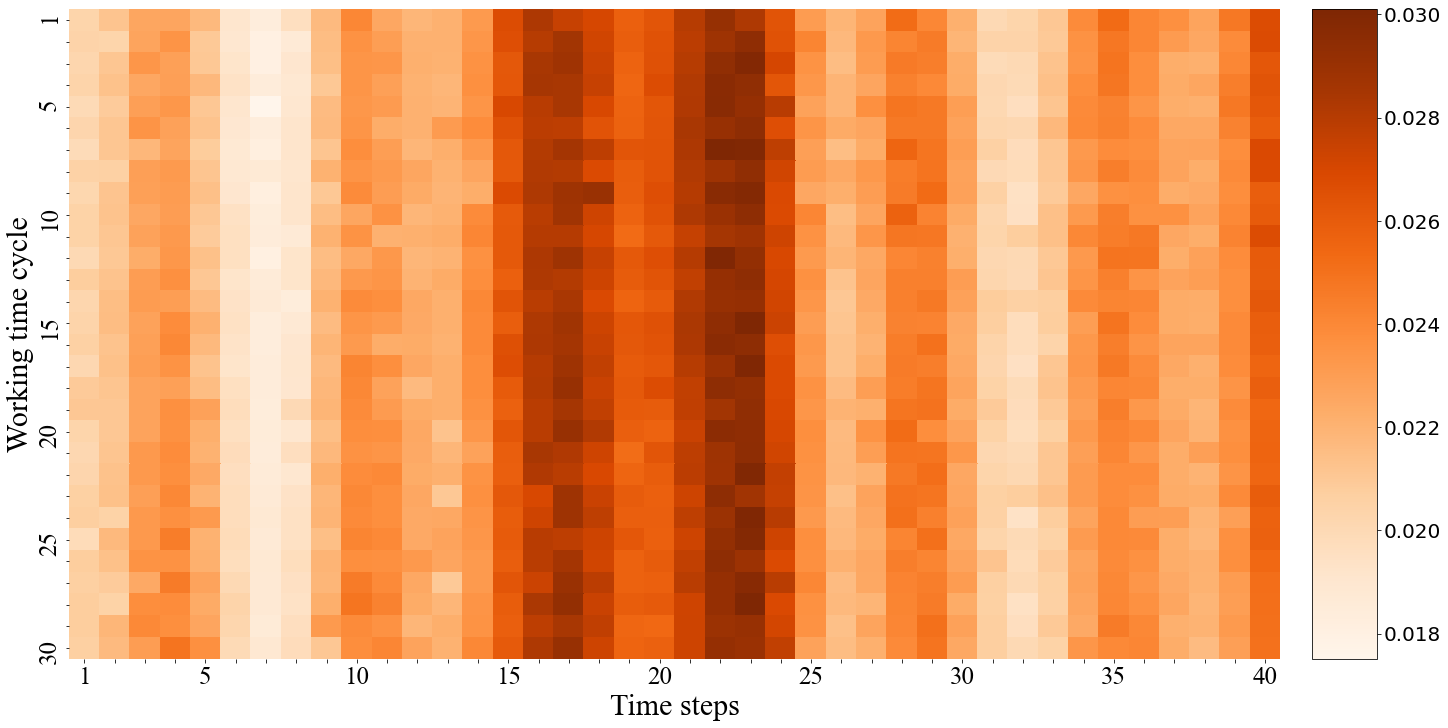}}
\caption{The average weights of each sensor and time step.}
\label{feature map}
\end{figure}

One advantage of DAST is that, the two encoders can automatically learn the weights of different sensors and time steps, which represents their importance for the real-time RUL prediction. The weight information is not only useful in improving the prediction performance, but also can be understood by the maintenance personnel to focus on more important sensors and time steps in real time and thus improve maintenance efficiency. To visualize this point, we choose 30 consecutive working time cycles (150-180) of engine unit 99 in the F003 dataset, and plot the average weights of each sensor and time step in the corresponding sliding window in Fig. \ref{feature map}. We can see that in this period, sensors T50, Nc, and phi are more important than other sensors, while for time steps, those from 15 to 24 and the last step are more informative for RUL prediction.

\section{Discussion}
We highlight the advantages of the proposed method compared to the shortcomings of current methods in this section. First of all, existing deep learning based RUL prediction methods are mainly based on the RNN/CNN architecture \cite{wang2018remaining,li2018remaining,cheng2020remaining,kong2019convolution,zheng2017long}. Some recent studies tried to combine the RNN/CNN architecture and the Attention mechanism to propose the RUL prediction method \cite{liu2020remaining,xiang2020lstm,song2020distributed,chen2020machine}, which first extracts the feature information of sequence data through RNN/CNN, and then learns the importance of the feature information through the attention mechanism.
Although the above methods achieved relatively good prediction effect in RUL prediction, there are still many shortcomings. For one thing, due to the existence of the structure of RNN/CNN, the RUL prediction model still has a bottleneck in extracting long-term dependency information. For another, the problems in the RUL prediction methods built by combining the RNN/CNN architecture with the attention mechanism are that the input data enters the attention modules and RNN/CNN modules sequentially, which causes the mutual influence between the extracted different feature information, thereby affecting the RUL prediction performance.

Different from the RUL prediction method \cite{liu2020remaining,xiang2020lstm,song2020distributed,chen2020machine}, our method is built upon the Transformer architecture, which is purely based on the self-attention mechanism to process all CBM data points in the sequence without considering their distance. In particular, we propose the design of the duel aspect feature extraction in parallel, which not only improves the ability of the model in capturing long-term dependency information but also enables it to focus on the more important sensor and time step information respectively, so as to avoid the mutual influence of the two aspect information. Our experimental comparison with the above RUL prediction method have verified the advantage of our method. In the ablation experiments, we have verified the effectiveness of our parallel feature extraction design, which is significantly better than the vanilla Transformer structure \cite{vaswani2017attention}.

\section{Conclusions and Future Work}
In this paper, we propose a novel deep RUL prediction method named Dual Aspect Self-attention based on Transformer (DAST). Without any RNN/CNN structure, DAST uses the self-attention mechanism to process the entire CBM data sequence. In particular, it is designed based on a parallel feature extraction scheme, which employs a sensor encoder and a time step encoder to capture the weighted features of different sensors and time steps simultaneously. The proposed parallel encoding architecture which runs the two encoders simultaneously and then fuse the two set of features can avoid mutual influence of information from the two aspects. Without the need of human intervention, the DAST model can adaptively learn the importance of different sensors and time steps, which could be helpful to the maintenance personnel to focus on those more important sensors and time steps, so as to improve maintenance efficiency. We conduct ablation studies to prove the effectiveness of our design. Experimental results on two real turbofan engine datasets show that the RUL prediction performance of our method is superior to state-of-the-art deep RUL prediction methods. In the future, we would like to explore the combination of deep learning methods and data enhancement techniques to resolve the issue of lacking training data, which is important for practical RUL prediction.


\bibliographystyle{IEEEtran}
\bibliography{bibfile}

\begin{IEEEbiography}[{\includegraphics[width=1in,height=1.25in,clip,keepaspectratio]{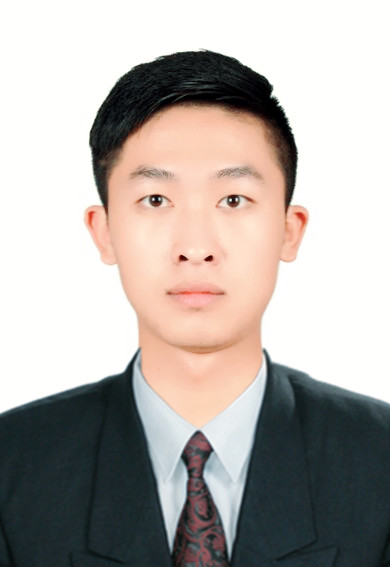}}]
{Zhizheng Zhang} received the M.S. degree in marine engineering of in Dalian Maritime University, Dalian, China, in 2020. He is currently pursuing the Ph.D. degree in control science
and engineering from Shandong University. His current research interests include deep learning, deep reinforcement learning, remaining useful life prediction, multivariate time series prediction.
\end{IEEEbiography}

\begin{IEEEbiography}[{\includegraphics[width=1in,height=1.25in,clip,keepaspectratio]{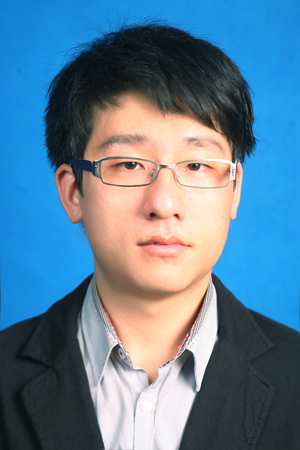}}]
{Wen Song}received the B.S. degree in automation and the M.S. degree in control science and engineering from Shandong University, Jinan, China, in 2011 and 2014, respectively, and the Ph.D. degree in computer science from the Nanyang Technological University, Singapore, in 2018. He was a Research Fellow with the Singtel Cognitive and Artificial Intelligence Lab for Enterprises (SCALE@NTU). He is currently an Associate Research Fellow with the Institute of Marine Science and Technology, Shandong University. His current research interests include artificial intelligence, deep reinforcement learning, planning and scheduling, and operations research.
\end{IEEEbiography}

\begin{IEEEbiography}[{\includegraphics[width=1in,height=1.25in,clip,keepaspectratio]{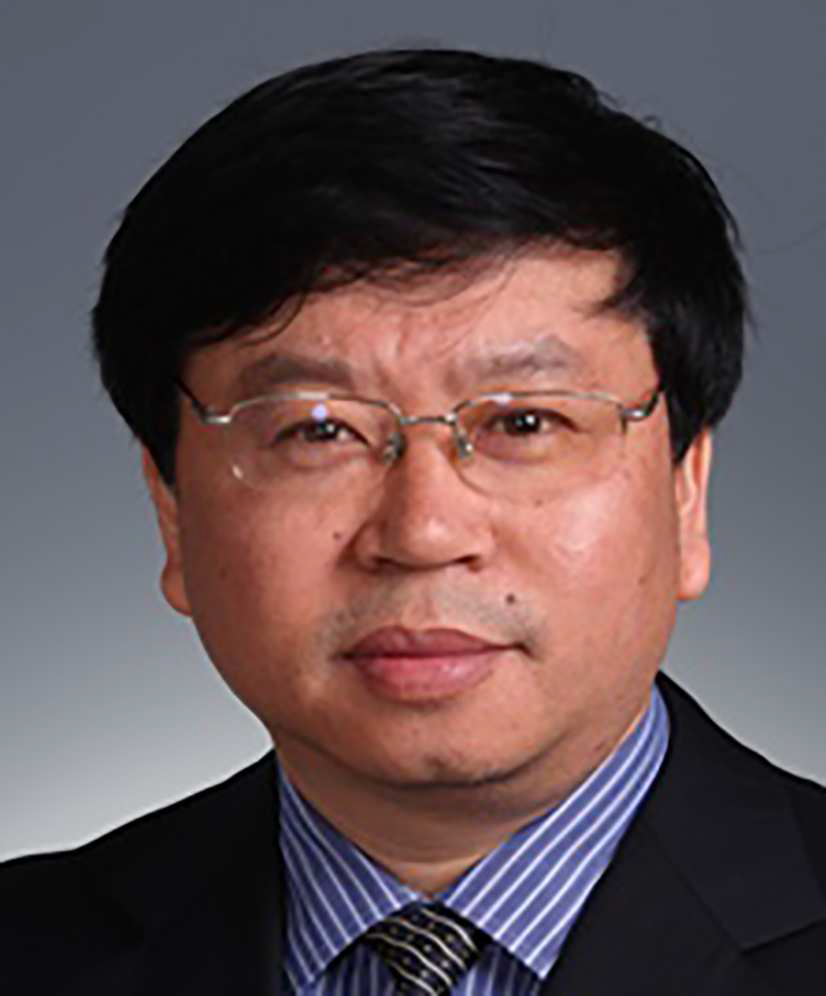}}]
{Qiqiang Li} received the Ph.D. degree in industrial automation from the Institute of Industrial Process Control, Zhejiang University in 1998. He is a Professor with the School of Control Science
and Engineering and the Institute of Marine Science and Technology, Shandong University. His research area focuses on modeling, optimization,
and simulation of complex systems. His current research interests are concerned with economic operation optimization of photovoltaic systems, energy efficiency of process
industry and commercial buildings.
\end{IEEEbiography}

\end{document}